\documentclass[compsoc,conference,a4paper,10pt,times]{IEEEtran}
\IEEEoverridecommandlockouts

\usepackage{listings}
\usepackage{xcolor}
 
\definecolor{codegreen}{rgb}{0,0.6,0}
\definecolor{codegray}{rgb}{0.5,0.5,0.5}
\definecolor{codepurple}{rgb}{0.58,0,0.82}
\definecolor{backcolour}{rgb}{0.95,0.95,0.92}
 
\lstdefinestyle{mystyle}{
    commentstyle=\color{codegray},
    numberstyle=\tiny\color{codegray},
    stringstyle=\color{codepurple},
     basicstyle=\ttfamily\footnotesize,
    breakatwhitespace=false,         
    breaklines=true,                 
    captionpos=b,                    
    keepspaces=true,                 
    numbers=left,                    
    numbersep=5pt,                  
    showspaces=false,                
    showstringspaces=false,
    showtabs=false,                  
    tabsize=2
}
\usepackage{cite}
\usepackage{amsmath,amssymb,amsfonts}
\usepackage{algorithmic}
\usepackage{graphicx}
\usepackage{textcomp}
\usepackage{bmpsize}
\usepackage{xcolor}
\usepackage{lipsum}
\usepackage{comment}
\usepackage[colorlinks=true,urlcolor=black]{hyperref}
\def\BibTeX{{\rm B\kern-.05em{\sc i\kern-.025em b}\kern-.08em
    T\kern-.1667em\lower.7ex\hbox{E}\kern-.125emX}}

\begin{document}

\title{A Way Around UMIP and Descriptor-Table Exiting via TSX-based Side-Channel}


\author{Mohammad Sina Karvandi$^{\ast}$,  Saleh Khalaj Monfared$^{\ast}$, Mohammad Sina Kiarostami$^{\ast}$, \\Dara Rahmati$^{ \ast}$, Saeid Gorgin$^{\ddag \ast}$\\
$^{*}$School of Computer Sciences, Institute for Research in Fundamental Sciences (IPM), Tehran, Iran\\ 
$^{\ddag}$Iranian Research Organization for Science and Technology (IROST), Tehran, Iran\\
\{karvandi, monfared, skiarostami, dara.rahmati\}@ipm.ir, gorgin@irost.ir }

\maketitle



\begin{abstract}

Nowadays, in operating systems, numerous protection mechanisms prevent or limit the user-mode applications to access the kernel's internal information. This is regularly carried out by software-based defenses such as Address Space Layout Randomization (ASLR) and Kernel ASLR (KASLR). They play pronounced roles when the security of sandboxed applications such as Web-browser are considered. Armed with arbitrary write access in the kernel memory, if these protections are bypassed, an adversary could find a suitable \textit{where to write} in order to get an elevation of privilege or code execution in \textit{ring 0}.
\par In this paper, we introduce a reliable method based on Transactional Synchronization Extensions (TSX) side-channel leakage to reveal the address of the Global Descriptor Table (GDT) and Interrupt Descriptor Table (IDT). We indicate that by detecting these addresses, one could execute instructions to sidestep the Intel's User-Mode Instruction Prevention (UMIP) and the Hypervisor-based mitigation and, consequently, neutralized them. The introduced method is successfully performed after the most recent patches for \textit{Meltdown} and \textit{Spectre}. Moreover, the implementation of the proposed approach on different platforms, including the latest releases of Microsoft Windows, Linux, and, Mac OSX with the latest $9^{th}$ generation of Intel processors, shows that the proposed mechanism is independent from the Operating System implementation. We demonstrate that a combination of this method with call-gate mechanism (available in modern processors) in a chain of events will eventually lead to a system compromise despite the limitations of a super-secure sandboxed environment in the presence of Windows's proprietary Virtualization Based Security (VBS). Finally, we suggest the software-based mitigation to avoid these issues with an acceptable overhead cost.

\end{abstract}


\section{Introduction}




As signs of progress in computer science, from Artificial Intelligence \cite{8848043} to High-Performance Computing \cite{hajihassani2019fast} continues, the role of computer security in both hardware and software is drawing more attention to the research community.
Recently discovered microarchitectural vulnerabilities in modern CPUs,  are known to be devastating. They are easy-to-implement, practical, and almost independent from the operating system, which makes them an imminent threat to computer privacy. Among them, speculative-execution based and side-channel attacks are more ubiquitous as new disclosures continue to increase scrutiny by researchers in this field\cite{ge2018survey}. These attacks are capable of circumventing all existing protective measures, such as CPU microcode patches, kernel address space isolation (Kernel Virtual Address (KVA), shadowing, and Kernel Page-Table Isolation (KPTI)). While side-channel attacks have been well-known for a relatively long time, speculative-execution based attacks are contemporary, and pieces of evidence indicate that they will persist for some time in the future.

 \par Pioneered by Meltdown \cite{lipp2018meltdown} and Spectre \cite{kocher2018spectre} attacks, numerous variations, and extension of microarchitecture vulnerabilities have been found, and their corresponding exploitation has proposed latterly. ForeShadow \cite{weisse2018foreshadow}, MDS \cite{minkin2019fallout}, and ZombieLoad \cite{schwarz2019zombieload} should be alluded as the most famous ones. Moreover, new works have shown the extensiveness of these attacks. As an example, NetCAT \cite{kurth_netcat} presents a practical network-based side-channel attack.

\par After Meltdown, more strict KASLRs such as  KAISER \cite{gruss2017kaslr} have been employed in today's operating systems to prevent similar attacks since short-term hardware mitigation is not effortlessly attainable. KAISER completely isolates the user-mode and kernel-mode memory layout by creating a \textit{Shadow} representation of the mapped memory. However, there are still some unprotected addresses and parts by KALSR that required by the architecture. Hence,  knowing these structure's addresses could lead to severe problems.

In addition, discovered hardware-based vulnerabilities on Memory (DRAM) such as RowHammer \cite{kim2014flipping} allow attackers to execute more destructive and offensive malicious code, to trespass or gain access to restricted and private information \cite{seaborn2015exploiting}.

Furthermore, it is possible and suitable to take advantage of some hardware-specific structures that are undoubtedly implemented across operating systems. In the same way, to gather masked and hidden internal information of the operating system could be used for malicious purposes. To be more precise, the structures of Global Descriptor Table (GDT) and Interrupt Descriptor Table (IDT) are one of the essential parts of protected mode, which are not heavily isolated in the user-mode and kernel-mode address layout. By overwriting these structures in certain conditions, one can perform a privilege escalation attack. Also, by the use of the same variations of timing side-channel attacks as in Meltdown,( e.g., TSX-based attacks), the virtual addresses of these structures in the kernel memory could be revealed.

In this work, we demonstrate that GDT and IDT addresses could be discovered by TSX side-channel to perform privilege escalation attacks, even after Meltdown mitigation, bypassing the mitigations in modern Intel processors, particularly User-Mode Instruction Prevention (UMIP). Furthermore, it is illustrated that the proposed attacks can be executed in virtualized environments, such as the latest Microsoft Hypervisor release (Hyper-v) and Virtualization Based Security (VBS). In summary, the contributions of this paper are as follow:
\begin{itemize}
    \item A TSX side-channel attack is performed to discover \textit{GDT} and \textit{IDT} addresses in the kernel mode in a system with \textit{KAISER} isolated memory layout bypassing UMIP.
    \item A full system compromise could be achieved by revealing \textit{GDT} and \textit{IDT} virtual addresses in the memory, incorporated with \textit{call-gate} mechanism along with a conventional \textit{Write What Where}.
    \item The possible mitigation investigated for this vulnerability and low-cost software-based mitigation for the operating systems to avert these attacks is suggested.
\end{itemize}

In the rest of the article, first, the necessary background information for the proposed attack, including a study on KASLR, Meltdown attack,\textit{Virtualization Base Security}, \textit{KAISER}, and other related concepts is provided in Section 2. GDT and call-gate Mechanism are explained in detail in section 3. In the section 4, Intel's UMIP is analyzed and described. Section 5 presents the attack implementation and experimental details by taking advantage of some exploitation methods. Possible mitigation for this vulnerability is discussed in section 6. Finally, other related works are noted and briefly investigated in section 7. The paper is summarized and concluded in section 8.

\section{Background}

In this section, required preliminaries and background for the software-based side-channel attacks, address space switching, along with some concepts of Translation Lookaside Buffer (TLB), Virtualization Based Security (VBS) have been provided. Moreover, additional materials on VM-Execution Controls, KAISER, Virtual Machine Control Structure (VMCS), TSX side channels, and Descriptor-Table Exiting are presented.

\subsection{KASLR and Meltdown}

The security of computer systems fundamentally relies on memory isolation, e.g., kernel address ranges are marked as non-accessible or, protected from user access. ASLR is a well-known technique to make exploitation harder by placing various objects randomly rather than using fixed addresses. It helps to ensure that memory addresses associated with running processes on systems are not predictable. Therefore, flaws or vulnerabilities associated with these processes will be more challenging to exploit.
Discovered Meltdown \cite{lipp2018meltdown} attack was able to exploit side effects of out-of-order execution on modern processors to read arbitrary kernel memory locations, including personal data and passwords. By exploiting the out-of-order execution as an indispensable performance feature, the attack is independent of the operating system, and it does not rely on any software vulnerabilities. Meltdown breaks all security guarantees provided by address space isolation as well as \textit{paravirtualized} environments and all of security mechanisms building upon this foundation. On the affected systems, Meltdown enables an adversary to read the memory of other processes or virtual machines in the cloud without any permissions or privileges, affecting millions of customers and virtually every user of a personal computer \cite{lipp2018meltdown}.

\subsection{Post Meltdown Patches}

Generally, Meltdown mitigation relies on isolating kernel and user memory pages with different methods. The widely used approach to address this issue is the employment of KAISER \cite{gruss2017kaslr}, which is implemented as Kernel Virtual Address Shadow (KVAS) (a term coined by Microsoft) \cite{kvas2018} in Microsoft Windows and KPTI in Linux \cite{grussKernelIsolation}.

Conventionally, before Meltdown, each process was equipped with a single set of page tables. KAISER \cite{gruss2017kaslr} proposes the implementation of two sets of page tables. One set is virtually unchanged and mapped when the process is in kernel mode. So, it includes both user-mode and kernel-mode memory sections. The second set (CR3) contains a copy of all of the user-space mappings but leaves out the kernel side. Instead, there is a small (minimal) set of kernel-space mappings that provides the minimum required information to the processor. This implementation of the dual page table prevents the adversary from gathering information regarding the kernel-space memory mapping scheme, avoiding further kernel-side exploitation. The concept of implementing KAISER is depicted in Figure \ref{KAISER} below:

\begin{figure}[h] 
\centering
\includegraphics[width=0.8\linewidth]{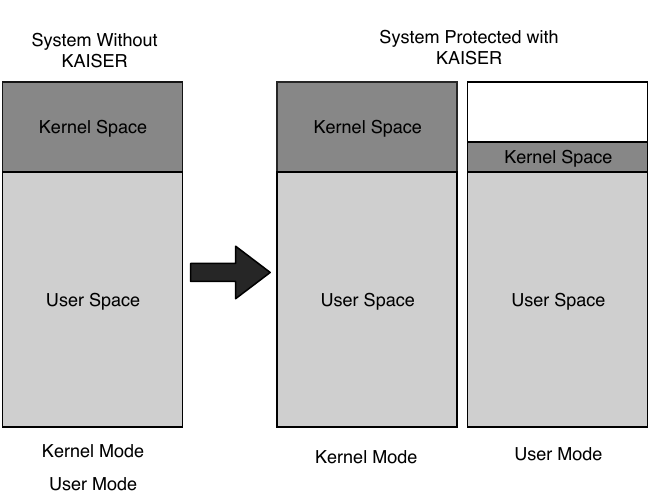}
\caption{KAISER protection overview before and after Meltdown Patch}
\label{KAISER}
\end{figure}

As shown in Figure \ref{KAISER}, placing a small portion of information in the user-mode is inevitable since operating systems are required to implement functions necessary to handle system calls and interrupts, which are directed to kernel space. Consequently, these shadowing functions change the base pointer for paging (e.g., \textit{CR3}) for a new page table.

A similar mechanism has been introduced and implemented in Microsoft‘s updates with regards to Meltdown based attacks in the KVAS. This feature effectively blocks the Meltdown attack, as it mislay a reasonably small portion of the kernel memory accessible to user-mode code.
In this system, the memory is partitioned into three parts: \textit{Entries}, \textit{Arbitrary Control Flow}, and \textit{Exits}. The key insight in this mechanism is that the kernel space and the user space are separated owing to advanced paging structures. Only minimal numbers of pages are mapped in both the user and the kernel spaces. As a result, even if a Meltdown attack is successful, it could not be used to leak kernel memory. That is due to swapping in address spaces by entries and exits, which leads to exclusive access to kernel space only by the kernel code. Another benefit of this design is to achieve its goal simply by manipulating the paging structures without having to rely on any extra support at the hardware level (e.g., microcode updates).

As will be discussed, leaving the tables which hold the addresses of interrupt handler (e.g., Interrupt Descriptor Table) or other tables managing the segmentation (e.g., GDT) visible to user mode, and ignoring to protect their addresses, allow the attacker to endanger the system. However, to adversely take advantage of the information left unprotected in the user-mode, essential internal mechanisms should be known which will be explored later.

\subsection {Address Space Switch}
On an address-space switch, as occurs on a process switch but not on a thread switch, some TLB entries can become invalid since the virtual-to-physical mapping is different. The most straightforward strategy to deal with this is to flush the TLB thoroughly. It means that after a switch, the TLB is empty, and any memory reference will be a miss, so it would be some time before things are running back at full speed. Newer CPUs use more effective strategies marking. It means that if a second process runs for only a short time and jumps back to a first process, it may still have valid entries, saving time to reload them.

\begin{figure}[!h] 
\centering
\includegraphics[width=0.8\linewidth]{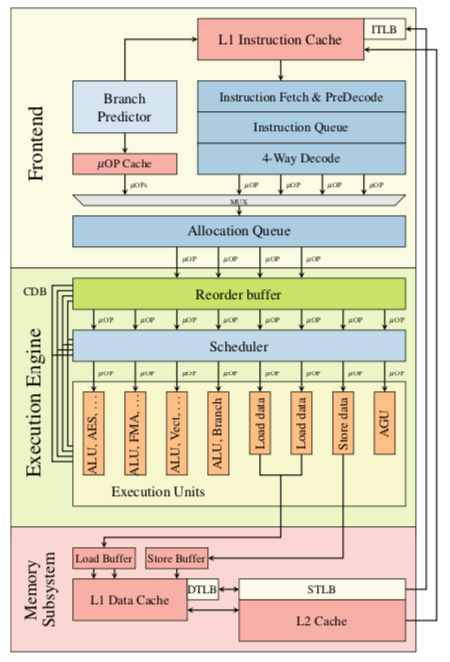}
\caption{Interconnection details in Intel SkyLake microarchitecture \cite{lipp2018meltdown}}
\label{tlbb}
\end{figure}

Since the 2010 Westmere microarchitecture Intel 64 processors also support 12-bit process-context identifiers (PCIDs)\cite{WestmereArrives}, which permit retaining TLB entries for multiple linear-address spaces, with only those that match the current PCID used for address translation.\cite{TLB_and_Meltdown}\cite{IntelPCID}
In Figure \ref{tlbb} the interconnection between different parts of caches like TLB (ITLB, STLB) and Shared Caches (L1, L2, L3) in Intel's SkyLake microarchitecture is illustrated.

While selective flushing of the TLB is an option in software-managed TLBs, the only option in some hardware TLBs (e.g., the TLB in the prior Intel processors) is the complete flushing of the TLB on an address-space switch.\cite{GilTenePCID}

Memory isolation is especially critical during switches between the privileged operating system kernel process and the user processes as was highlighted by the Meltdown security vulnerability. Mitigation strategies such as KPTI rely heavily on performance-impacting TLB flushes and benefit significantly from hardware-enabled selective TLB entry management such as PCID.

\subsection{Virtualization Base Security (VBS)}

\par Virtualization-based security, or VBS, uses hardware virtualization features to create and isolate a secure region of memory from the standard operating system. Windows can use the \textit{virtual secure mode} to host several security solutions, providing them with significantly increased protection from vulnerabilities in the operating system and preventing the use of malicious exploits that attempt to defeat protections.

\par VBS uses the Windows hypervisor to create secure virtual mode and to enforce restrictions which protect the vital system and operating system resources, or to protect security assets such as authenticated user credentials. With the increased protections offered by VBS, even if malware gains access to the operating system kernel, the possible exploits can be notably limited and contained, because the hypervisor can prevent the malware from executing code or accessing platform secrets.

\par One such example security solution is Hypervisor-Enforced Code Integrity (HVCI) \cite{hvci}, which uses VBS to strengthen code integrity policy enforcement significantly. Kernel-mode code integrity checks all kernel-mode drivers and binaries before they started, and prevents unsigned drivers or system files from being loaded into system memory.
The presence of this feature can mitigate the execution of LGDT, LIDT, LLDT, LTR, SGDT, SIDT, SLDT, and STR, so GDT and SDT not included to address leak to either operating system kernel-mode (ring 0) and user-mode (ring 3). These leaks can be prevented by using the second bit of Secondary Processor-Based VM-Execution Controls.

\subsection{Secondary Processor-Based VM-Execution Controls}

In order to control our guest features, we have to set some fields in our Virtual Machine Control Structure (VMCS), which is a hardware-defined structure that controls the behavior and settings of each guest virtual machine (VM).

This data structure is located in memory and exists once per (current) VM, which is managed by the Virtual Machine Monitor (VMM). With every change of the execution context between different VMs, the VMCS is restored for the current VM, defining the state of the VM’s virtual processor and VMM control Guest software using VMCS.

The VMCS consists of six logical groups:

\begin{itemize}
\item \textbf{Guest-State Area}: Processor state saved into the guest state area on VM exits and loaded on VM entries.
 \item \textbf{Host-State Area}: Processor state loaded from the host state area on VM exits.
\item \textbf{VM-Execution Control Fields}: Fields controlling processor operation in VMX non-root operation.
\item \textbf{ VM-Exit Control Fields}: Fields that control VM exits.
\item \textbf{ VM-Entry Control Fields}: Fields that control VM entries.
\item \textbf{ VM-Exit Information Fields}: Read-only fields to receive information on VM exits describing the cause and the nature of the VM exit.
\end{itemize}

Secondary Processor-Based VM-Execution Controls \cite{secondary_vmbased2019intel}, which is a member of VMCS along with Primary Processor-Based VM-Execution Controls fields\cite{primary_procbased2019intel} control these features that can be modified using VMWRITE instruction. Several features described above control the presence and absence of sundry instructions and security mechanisms and behavior of guests when, for example, a particular instruction is executed. Among the mentioned features, Descriptor-Table Exiting is considered in this work. If this control bit is set, then the guest is no longer able to execute instructions such as LGDT, LIDT, LLDT, LTR, SGDT, SIDT, SLDT, and STR directly into VMX non-root mode \cite{SinaKarvandi2018HVFS_Part1}. In this situation, instead, a VM-Exit occurs, and then, it is the responsibility of VMM to handle the results to the guest. The VMM could decide whether to return a valid or invalid result to the guest, accordingly. This controlling feature could be used as a mitigation to avoid these data leaks to the user-mode or kernel-mode. In the suggested  scenario, we show that our proposed attack is independent of the returned results from SGDT, SIDT, SLDT, and STR, and would not cause a VM-exit \cite{SinaKarvandi2018HVFS_Part5}.


\subsection {Integrity Levels in Windows}

Beginning with Windows Vista operating system, the Windows integrity mechanism improved the security architecture by defining a new access control entry (ACE) type to represent an integrity level in an object's security descriptor.

The security descriptor is a data structure containing the security information associated with a \textit{securable object}.
In Windows, contrary to Linux, one could read kernel addresses using a popular function called \textit{NtQuerySystemInformation}. It is due to the fact that, in Windows, KASLR is not a boundary against local attackers with unconstrained execution. Therefore,  it is meaningless if an adversary application is executed in Integrity Level (which is equally or more protected than Medium Level). despite of the multiple options to exploit Windows kernel, Integrity Levels are designed to prevent untrusted sources from accessing kernel addresses. For example, applications with \textit{Low} or \textit{Untrusted} integrity levels such as Web-browsers are prevented from reading these addresses.
 
This is a defense mechanism in fundamental design  to protect operating system kernel, and it is imperative to protect other applications with different levels of trust to be isolated from each other (e.g., User Interface Privilege Isolation (UIPI) \cite{uipi}). In our attack senario, we demonstrate how to find the address of GDT from a low-integrity application. In our representation, we inject our work (DLL) into the Microsoft Edge (bypassing some DLL-injection protections) to test our samples. Furthermore, There are no similar mechanism to integrity-level in other operating systems. They instead use account-level policies to restrict a malicious application from affecting other parts of applications.

\subsection{Descriptor-Table Exiting}

Descriptor-Table Exiting is a hardware mechanism to restrict guest machines in VMX Non-Root from executing instructions such as LGDT, LIDT, LLDT, LTR, SGDT, SIDT, SLDT, and STR. 
This mechanism has been used in Microsoft Virtualization Based Security as exploit mitigation, which avoids memory address leakage and provides an absurd situation for the attacker to find the base address of GDT or IDT, among other details such as Control Registers. This outcome is because Microsoft uses hypervisor as a hardware security mechanism, and in VM Control Structure, there is a field for configuring this hardware feature, called \textit{Descriptor-Table Exiting}.

Descriptor-Table Exiting is declared in Intel Manual \cite{primary_procbased2019intel}.
This control determines whether executions of LGDT, LIDT, LLDT, LTR, SGDT, SIDT, SLDT, and STR cause VM exits. This declaration would be critical to the attack model we intend to describe.

\subsection{TSX Cache Attack}

By the use of Intel TSX, which is a product name for two x86 instruction set extensions, called Hardware Lock Elision (HLE) and Restricted Transactional Memory (RTM).\cite{stecklina2018lazyfp}, the initial phase of the attack is triggered.
HLE is a set of prefixes that could be added to specific instructions. These prefixes are backward-compatible. Hence, the code, including them, also works on older hardware platforms.
On the other hand, RTM is an extension adding several instructions to the instruction set that are used to declare regions of code that should execute as part of a hardware transaction. Transactions can protect a series of memory accesses that shall either all succeed together or shall be rolled back together in case of any error condition or concurrent access by other threads.

\par A RTM transaction comprises the region of the code that is encapsulated between a pair of \textit{xbegin} and \textit{xend} instructions. Instruction xbegin also provides a mechanism to define a fall-back handler that is called if the transaction is aborted and \textit{xabort} can be used by the executing code to abort the transaction explicitly. Besides, the processor might abort the transaction upon certain events. These events include an exception that occurs during the transaction. In this paper, by referring to Intel TSX, we expect RTM specifically. TSX is vital in terms of security as it used in many side-channel attacks, and it makes timing side-channels more precise by handling errors in the transaction failed section (in user-mode).

By employment of the TSX, generating an exception or an interrupt which is handled in the kernel could be avoided, resulting in side-channel attacks more resistant to noise and improvement in outcomes.

\section{Attacks Based on GDT Access}

As an indispensable part of the suggested attack, GDT and its properties are described in detail in the this section.

\subsection{Global Descriptor Table}

\begin{figure}[h] 
\centering
\includegraphics[width=0.95\linewidth]{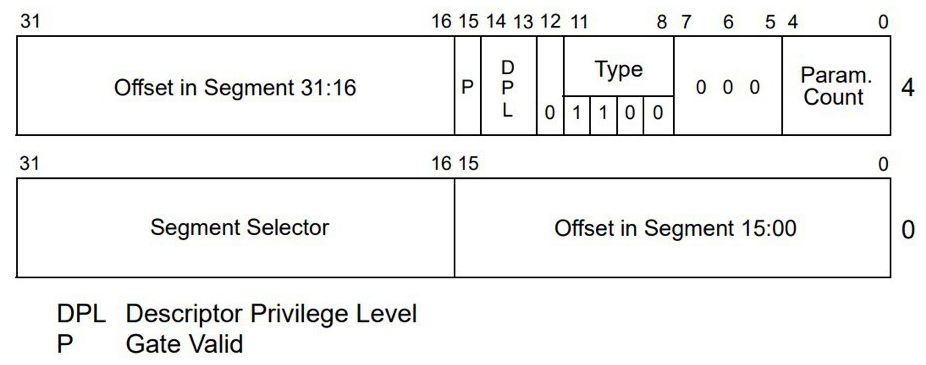}
\caption{GDT structure in a 32-bit machine}
\label{GDT}
\end{figure}

GDT is a data structure employed by Intel x86-family processors starting with the 80286 in order to define the characteristics of the various memory areas used during program execution, including the base address, the size, and access privileges such as executability and writability. GDT is a main table in x86 and protected-mode that still exists in AMD64 \cite{devices2006amd64} and Intel IA-32e. The GDT structure in the x86 system is shown in Figure \ref{GDT}.

While the proposed attack here works on both x86 and x64 architectures, we have used the x64 version of GDT since it is more widespread rather than the other version.

\subsection{GDT in 64-bit}

In the modern systems in protected-mode with paging enabled, although the segmentation is omitted, the GDT still presents in 64-bit mode. A GDT must be defined but is generally never changed or used for segmentation. The size of the register has been extended from 48 to 80 bits, and 64-bit selectors are always \textit{Flat} (thus, from \texttt{0000000000000000} to \texttt{FFFFFFFFFFFFFFFF}). However, the base of FS and GS are not constrained to 0, and they proceed to be used as pointers to the offset of items such as the process environment block and the thread information block e.g., in x86 version of Windows FS points to \textit{\_TEB} structure for the current thread in user-mode, and \textit{\_KPRCB} in kernel-mode and GS have the same usage in x64 machines.

64-bit versions of Microsoft Windows forbid hooking of the GDT. Attempting to do so would cause the machine to \textit{bug check}. It is not a problem for our case as long as mechanisms for preventing these hooks called Kernel Patch Protection, which is known as \textit{PatchGuard}, checkd the system in random intervals of between 3 to 10 minutes. So we can patch GDT in a glance then make everything back again to avoid such errors. In this context, we use GDT as a descriptor for call-gate to complete the attack chain instead of a descriptor for segmentation. 

\begin{figure}[h]
\centering
\includegraphics[width=0.95\linewidth]{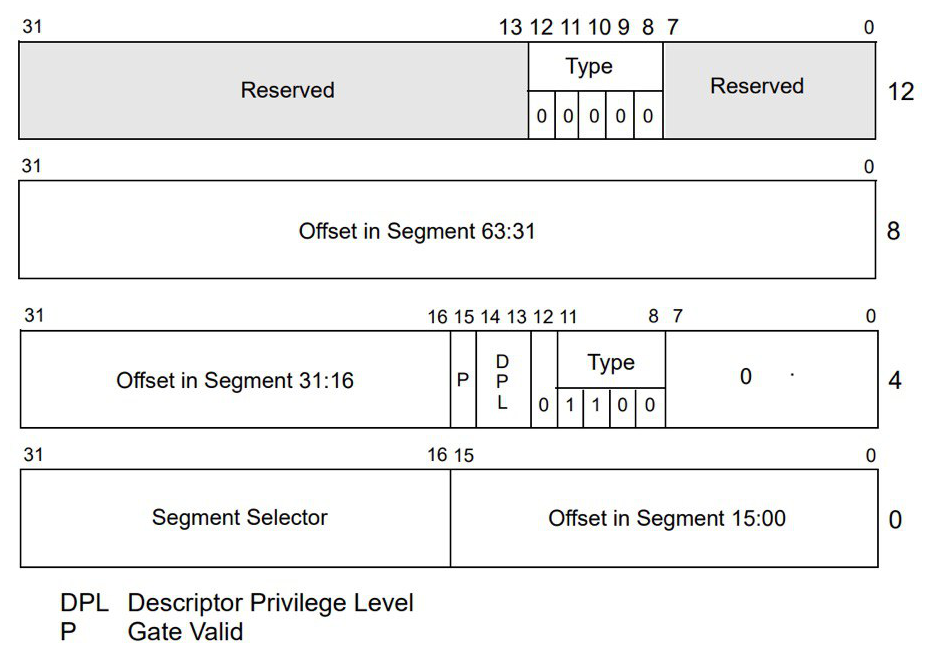}
\caption{GDT Structure in x64}
\label{lfsr}
\end{figure}

\subsection{Call-gate Mechanism}

Call-gates are used to transfer the execution to other rings e.g., ring 0, 1, 2, 3. Instructions like \textit{SYSENTER} and \textit{SYSCALL} are used in modern operating systems for transitioning between every rings to ring 0. But for the transition between other rings (e.g., ring 3 to 2 or 2 to 1), the call-gates would be used.
The \textit{type} field located in the GDT structure as indicated in Figure \ref{GDT} represents a 4-bit field that could get various values and completely change the GDT entry behavior and definition. So, it could be filled with one of the values indicated in Table \ref{tab:1} depending on the entry's usage\cite{SinaKarvandi2019CallGates}.
\begin{table}[h!]
\centering
\caption{The Possible Values for Type field in GDT}

\begin{tabular}{|c|c|}
\hline
0000 & Reserved                     \\ \hline
0001 & Available 16-bit TSS         \\ \hline
0010 & Local Descriptor Table (LDT) \\ \hline
0011 & Busy 16-bit TSS              \\ \hline
0100 & 16-bit call-gate             \\ \hline
0101 & Task Gate                    \\ \hline
0110 & 16-bit Interrupt Gate        \\ \hline
0111 & 16-bit Trap Gate             \\ \hline
1000 & Reserved                     \\ \hline
1001 & Available 32-bit TSS         \\ \hline
1010 & Reserved                     \\ \hline
1011 & Busy 32-bit TSS              \\ \hline
1100 & 32-bit call-gate             \\ \hline
1101 & Reserved                     \\ \hline
1110 & 32-bit Interrupt Gate        \\ \hline
1111 & 32-bit Trap Gate             \\ \hline
\end{tabular}
\vspace{3mm}
\label{tab:1}
\end{table}

As could be seen from the permissible values, after finding the target entry, the type  value should be changed to one \textit{Gate} accordingly. For example, we use 0xc (1100 - 32-bit call-gate) in the final payload.

There are some terms in call-gate used to build the final payload. In order to exploit the features that call-gate provides, the suitable privilege level should be set in the data segmentation used in the GDT. Here are the privilege levels defined in this context.
\begin{itemize}

\item \textbf{Current Privilege Level (CPL)}
CPL is stored in the selector of currently executing the CS register. It represents the privilege level (PL) of the currently executing task and also PL in the descriptor of the code segment and designated as Task Privilege Level (TPL) \cite{SinaKarvandi2019CallGates}.

\item \textbf{Descriptor Privilege Level (DPL)}
It is PL of the object which is being attempted to be accessed by the current task or put differently, the least privilege level for the caller to use this gate \cite{SinaKarvandi2019CallGates}.

\item \textbf{Requester Privilege Level (RPL)}
It is the lowest two bits of any selector. It can be used to weaken the CPL if craved \cite{SinaKarvandi2019CallGates}.

\item \textbf{Effective Privilege Level (EPL)} It is maximum of CPL and RPL thus the task becomes less privileged \cite{SinaKarvandi2019CallGates}.
\end{itemize}

\par It is assumed that a task needs data from the data segment. Therefore, the privilege levels are checked at the time a selector for the target segment is loaded into the data segment register. Three privilege levels enter into the privilege checking mechanism.
Ultimately, the payload must meet the following conditions in the fields:
\begin{itemize}
    
 \item RPL of the selector of the target segment.

 \item DPL of the descriptor of the target segment
\end{itemize}
Note that the access is allowed only if \textit{DPL} is greater than or equal to the maximum of CPL and RPL, and a procedure can only access the data that is at the same or less privilege level.

\subsection{From call-gate to code Execution in Ring-0}\label{AA}
\subsubsection{Call-gate in x86}
In order to use x86, fields of a unique set of call-gate should be filled as described in Table \ref{tab:2}.

Selector field should be 0x8 to point to KGDT\_R0\_CODE entry of GDT, which describes the kernel-mode in Windows. The type of it should be 0xc, and the minimum ring that can invoke this call-gate is 0x3 (DPL = 0x3 (user-mode)), and also, it should be present in memory (pFlag = 0x1) \cite{SinaKarvandi2019CallGates}.
\begin{table}[h!]
\caption{Organization of the fields in the GDT}

\centering
\begin{tabular}{|c|c|}
\hline
selector      & 0x8                                                      \\ \hline
type          & 0xc                                                      \\ \hline
dpl           & 0x3                                                      \\ \hline
pFlag         & 0x1                                                      \\ \hline
offset 0\_15  & 0x0000ffff \& address                                    \\ \hline
offset 16\_31 & 0x0000ffff \& ( address \textgreater{}\textgreater{}16 ) \\ \hline
\end{tabular}
\vspace{3mm}
\label{tab:2}
\end{table}

\subsubsection{Call-gate in Long Mode}
Call-gate are unavoidable parts of Intel structure, and even in 64-bit long mode. In addition to GDT, LDT is also present but special cases like segmentation using the FS/GS segment are replaced by the new MSR-based mechanism using \textit{IA32\_GS\_BASE} and \textit{IA32\_KERNEL\_GS\_BASE} MSRs \cite{ia32_gs_base2019intel}.

The fact that LDT \& GDT are still presented in long mode is used in Windows when the kernel uses the UMS (User-Mode Scheduling), so Windows creates a Local Descriptor Table if a thread tends to use UMS \cite{alex20181}.

\section{User-Mode Instruction Prevention (UMIP)}
UMIP is a security feature present in new Intel Processors. If enabled, it prevents the execution of particular instructions if the Current Privilege Level (CPL) is greater than 0. If these instructions were executed when CPL $>$ 0, user space applications could have access to system-wide settings such as the global and local descriptor tables, the task register and the interrupt descriptor table. These are the instructions covered by UMIP:
\begin{itemize}

\item SGDT : Store Global Descriptor Table
\item SIDT : Store Interrupt Descriptor Table
\item SLDT : Store Local Descriptor Table
\item SMSW : Store Machine Status Word
\item STR : Store Task Register
\end{itemize}

If any of these instructions are executed with CPL $>$ 0, a general protection exception (GP) is issued when UMIP is enabled. In order to enable this feature, operating systems can set the $11^{th}$ bit of the CR4. In our observations, Linux and Windows do not use these features for some compatibility issues. Thus, this opens a kernel memory address leak to user-mode applications, and these valid addresses can be used for exploiting the Operating System Kernel or as a valid address for other side-channel measurements. In section 5, we demonstrate how these addresses could lead to a full system compromise.

Nevertheless, Microsoft decided to remove the support for GDT, SIDT, SLDT, SMSW, and STR instructions in hypervisor as explained. Our observation shows that even if operating systems use UMIP or DESCRIPTOR-TABLE EXITING separately or both of them simultaneously, it is still vulnerable to side-channel attacks based on TSX.

\subsection{Far Calls and Far JMPs}\label{BB}
The far forms of JMP and CALL refer to other segments and require privilege checking. The far JMP and CALL can be performed in two methods:
\begin{itemize}

\item Without call-gate Descriptor:
The processor permits a JMP or CALL directly to another segment only if:
\begin{enumerate}

\item DPL of the target segment = CPL of the calling segment

\item Confirming bit of the target code is set and DPL of the target segment $\leq$ CPL
\end{enumerate}

Note that Confirming Segment may be called from various privilege levels, but is executed at the privilege level of the calling procedure.

\item With call-gate Descriptor:
The far pointer of the control transfer instruction uses the selector part of the pointer and selects a gate. The selector and offset fields of a gate form a pointer to the entry of a procedure.
\end{itemize}

\section{Attack implementation}
In this section, we describe how the explored mechanism are used to create the attack. Then, we show the results obtained from the Intel processor and show how the valid base address of IDT and GDT could be obtained without using SIDT and SGDT. Next, we show how to build a valid call-gate entry and use it in combination with a write-what-where to execute an adversary code. Then attacker crafts the shellcode in \textit{ring 0} in order to elevate privilege or hide the malware in the kernel.
\subsection{Threat Model}
As a basic assumption for the attack model, the attacker can execute code in the victim’s computer in a limited level of privilege, including a highly limited user-mode or in a sandboxed application with all the common defenses (e.g., SMEP, SMAP, DEP) enabled and configured suitably. In order to fully compromise the system an attacker has prior \textit{write-what-where} (CWE-123)\cite{cwe123} vulnerability in operating system kernel. Further, as an extension to the proposed attack mechanism, the adversary might also execute code in a vitalized environment as well in the shared resource usage scenario.

\vspace{-2mm}
\subsection{Experimental Setup}
The experiment to showcase the effectiveness of the explained attack chain has been executed on a system equipped with  $9^{th}$ generation of Intel processor (i9-9880H), running on a Windows  19H1 (also known as 1903) with 16 GB of DDR4 RAM.

Moreover, the same attack procedure is carried out on a system with a $6^{th}$ generation CPU (6820HQ), to ensure the generalization of the method. The test has also been successfully experimented on 19H2 and the latest 20H1 Microsoft Windows, Ubuntu Debian 7, and Mac OSX Mojave as well.

\subsection{Finding GDT Address}
In order to locate the GDT address, a timing measurement is required to discover the elapsed time in accessing a \textit{mapped} and an \textit{unmapped} address in the kernel space memory.  Experimentally, a valid address gives the response time about 190 $\sim$ 197 clock-cycles (different based on architecture) and an invalid address access returns after about 220 $\sim$ 234 clock-cycles based on our results in $6^{th}$ Gen Intel (6820HQ).
\\
To implement such a measurement, a combination of the kernel memory address and access time (RDTSCP) + TSX (XBEGIN, XEND) is employed. Then the response time difference in accessing a mapped and unmapped addresses could lead to the identification of mapped addresses.

\lstset{numbers=left, numberstyle=\tiny, language=[x86masm]Assembler,  escapeinside={\%*}{*)},          
}
\begin{lstlisting}[caption={The timing measurement code deployed by the use of TSX technology (RDTSCP)}, label={lst:exemplo}, firstnumber=1, frame=single]
 %*\color{magenta}rdtscp *)  ; get the current time clock of processor
  ...       ; save the rdtscp results somewhere (e.g registers)

 %*\color{magenta}mov*) rax,[Kernel Address]  ; Move a kernel address into tax 
 %*\color{magenta}xbegin*)  $+xxx    ; Use Intel TSX in order to suppress any error in user-mode
 ; The error always happens because we are trying to read kernel address
 %*\color{magenta}mov byte*) ptr [rax], 0   ; Try to write into kernel address
  ...      ; Error occurs here (program never reaches here)

 %*\color{magenta}xend *)   ; End of TSX 
 %*\color{magenta}rdtscp *)  ; Compute the core clock timing again in order to see how many
        ; clocks CPU spends when trying too write into our address
\end{lstlisting}

Furthermore, if a particular processor does not support the RDTSCP instruction, then one could get similar results by the serialization process. More precisely, it is required to serialize instructions to execute all of the instructions fetched before the targeted instruction. So a combination of CPUID + RDTSC is adequately employed. Given the explained circumstances, the previous code could be modified as follows in Listing \ref{lst:exemplo1}:

\lstset{numbers=left, numberstyle=\tiny, language=[x86masm]Assembler,  escapeinside={\%*}{*)},          
}
\begin{lstlisting}[caption={The timing measurement code by serialization of instructions (RDTSC+CPUID)}, label={lst:exemplo1}, firstnumber=2, frame=single]
 %*\color{magenta} cpuid *)  ;Execute a serialization instruction
 %*\color{magenta}rdtsc *)  ; get the current time clock of processor
  ...       ; save the rdtsc results somewhere (e.g registers)

 %*\color{magenta}mov*) rax,[Kernel Address]  ; Move a kernel address into tax 
 %*\color{magenta}xbegin*)  $+xxx    ; Use Intel TSX in order to suppress any error in user-mode
                 ; The error always happens because we are trying to read kernel address
 %*\color{magenta}mov byte*) ptr [rax], 0   ; Try to write into kernel address
  ...      ; Error occurs here (program never reaches here)

 %*\color{magenta}xend *)   ; End of TSX 
 %*\color{magenta}rdtsc *)  ; Compute the core clock timing again in order to see how many
        ; clocks CPU spends when trying too write into our address
\end{lstlisting}

Note that the first implementation indeed gives more precise results compared to executing RDTSC. Our experiments show that it is not suitable to use CPUID for the second RDTSC as it takes several clocks-cycles.
\par Furthermore, it would be possible to use the timing thread, if a operating system prohibits the usage of RDTSC or RDTCSP \cite{inteltsd}, or intercepts the execution of CPUID using Intel VMX \cite{SinaKarvandi2019HVFS_Part6} or Intel FlexMigration \cite{IntelFlexMigration}. Timing threads could even have a higher resolution rather than RDTSC/RDTCSP on many processors \cite{schwarz2017malware}\cite{schwarz2017fantastic}.
By deploying these instructions, an automatic process is triggered to find valid targeted addresses.

\begin{figure}[!h]
\centering
\includegraphics[width=0.95\linewidth]{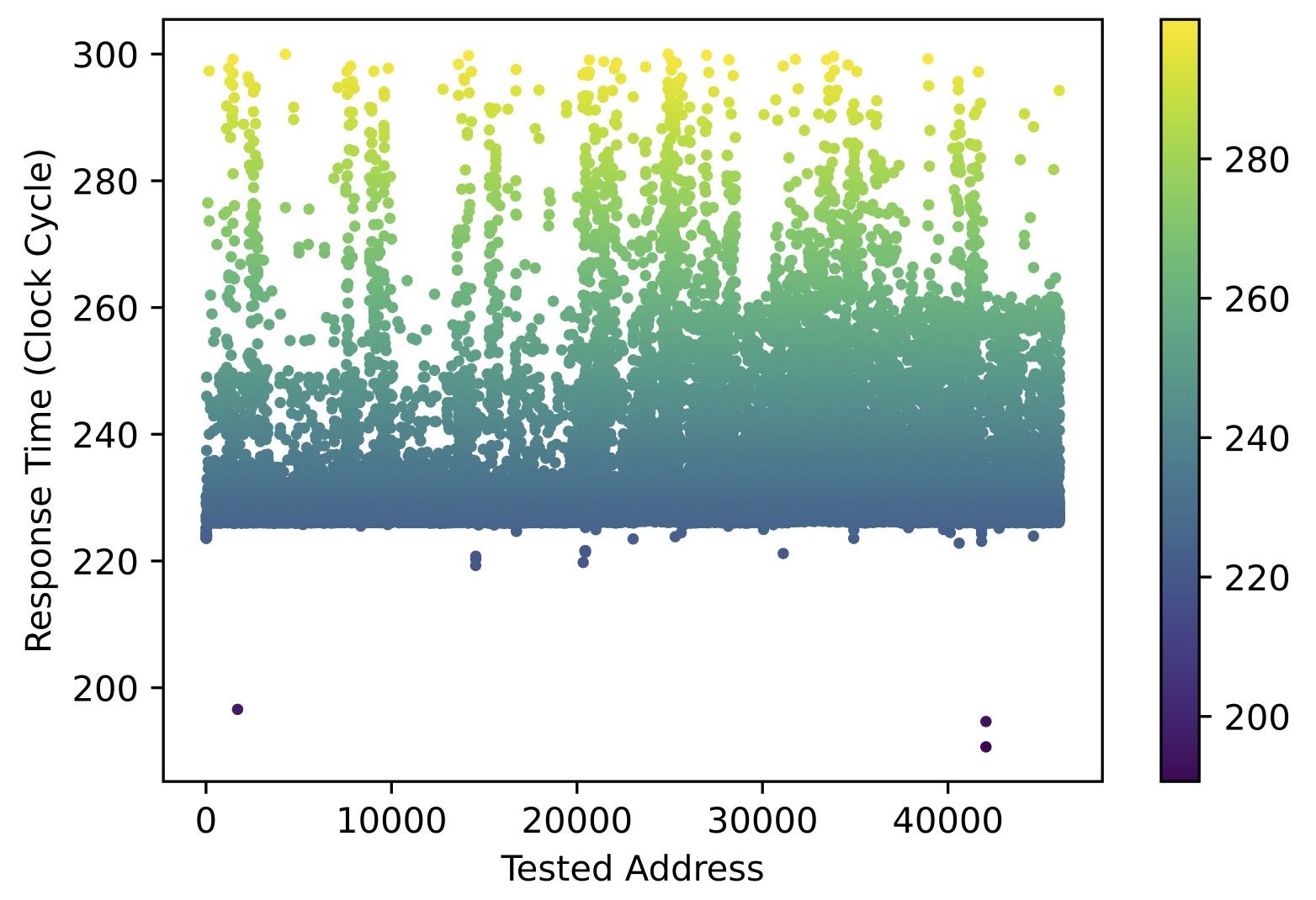}
\caption{The results of timing TSX-based measurements on a uni-core system}
\label{tsx}
\end{figure}

The result consists of four valid elements. The first one is the addresses that are valid for IDT. Second is the address of GDT, and third is the address of SYSCALL MSR\_LSTAR (\texttt{0xC0000082}) - (The kernel's RIP SYSCALL entry for 64-bit software)\cite{Osdev2017SysEnter}. Finally, the fourth is where the page tables are mapped. The timing results of the deployed measuring method is depicted in Figure \ref{tsx}.

Our observation in the latest 20H1 (and other versions of Windows) shows that GDT and IDT are mapped in a particular order, even though there is no limitation to assign different addresses.  By way of example, Windows maps IDT in a unique address. IDTR is \texttt{fffff80021eeb000}, and GDTR \texttt{fffff80021eedfb0} (GDTR = GDT Base + GDT size) and this sequence is the same each time Windows is restarted when the KASLR addresses changed. The difference is \texttt{0x2000} bytes or two pages. Thus, the address of IDT could first be determined, leading to revealing the address of GDT where another page of \texttt{0x2000} bytes is mapped following the first valid page address.

While there are other pages mapped into memory addresses (e.g., shadow functions for system-calls and interrupts), the addresses are far from the target addresses (e.g., \texttt{fffff8001d34e500}). Therefore, the address among IDT, GDT, Interrupt Shadows, and System Call Shadows could be identified. A payload for call-gate could build later finding the GDT base address.

The following commands in Listing \ref{lst:em4} shows the process of identification of valid addresses on all the cores by realizing the distance between GDT and IDT addresses.
\lstset{ numberstyle=\tiny, language=[x86masm]Assembler,  escapeinside={\%*}{*)},          
}
\begin{lstlisting}[caption={The procedure of employing IDTR and GDTR}, label={lst:em4}, firstnumber=2, frame=single]
 ; Accessing First Core's IDT and GDT
 0: kd> r %*\color{magenta}  idtr *) 
 %*\color{magenta} idtr=fffff8077925b000 *) 
 0: kd> r %*\color{magenta}  gdtr *)    
 %*\color{magenta} gdtr=fffff8077925dfb0 *)
 ; Accessing Second Core's IDT and GDT
 0: kd> %*$\sim$*)1 
 1: kd> r %*\color{magenta}  idtr *)   
 %*\color{magenta} idtr=ffff8401bc053000 *) 
 1: kd> r %*\color{magenta}  gdtr *) 
 %*\color{magenta} gdtr=ffff8401bc055fb0 *) 
 ; Accessing Third Core's IDT and GDT
 1: kd> %*$\sim$*)2
 2: kd> r %*\color{magenta}  idtr *)  
 %*\color{magenta} idtr=ffff8401bc0f5000 *)  
 2: kd> r %*\color{magenta}  gdtr *)   
 %*\color{magenta} gdtr=ffff8401bc0f7fb0 *)
 ; Accessing Forth Core's IDT and GDT
 2: kd> %*$\sim$*)3
 3: kd> r %*\color{magenta}  idtr *)  
 %*\color{magenta} idtr=ffff8401bc1a4000 *)  
 3: kd> r %*\color{magenta}  gdtr *)   
 %*\color{magenta} gdtr=ffff8401bc1a6fb0 *)  
\end{lstlisting}

We observed that allocated addresses for IDT and GDT have a special pattern for each core. For instance, here are several addresses that Windows allocated for IDT of 
its first core:
\begin{itemize}
    \item \texttt{fffff8036385b000}
    \item \texttt{fffff8027ca5b000}
    \item \texttt{fffff80053a5b000}
    \item \texttt{fffff8076525b000}
\end{itemize}
 Our experiments indicate that these addresses tend to follow a specific pattern. As the pseudo-code illustrated in Listing \ref{lst:em4}, the GDT has the same pattern As IDT as well. Our experiments show that, regardless of the system in hand, for the first core, the pattern of \texttt{fffff80XXXX5b000} is spotted, where \texttt{XXXX} can be changed due to the prevention mechanism of KASLR. The first bytes in the pattern address is to create a canonical address, and the least significant byte has a constant value of \texttt{5b000} pattern. This brings 0xffff = 65535 possibilities to find the address of IDT and GDT in the first core of Windows. The same pattern can be applied to other cores as well.
In a uni-core system, one can test up to 10 addresses per second with excellent precision, using the explained timing side-channel. Moreover, one could also hasten this measurement up to 20 addresses per second, in compromise to the loss of accuracy. Approximately, it takes 109 minutes to find the address of the GDT for the first core. Of course, the patterns for other cores could be discovered as well. As an example, in the 8-core system, there are eight possibilities for IDT and GDT addresses, which could speed up the search 8x faster. Also, it is possible to use other cores simultaneously for accelerating the search process.

\subsection{Build call-gate Entry}

We have built our payload based on the description discussed in section \ref{AA}.
\subsection{Using FAR JMPs, FAR CALLs}

As explored in section \ref{BB}, the near forms of JMP and CALL transfer within the current code segment requires only limited checking. However, the far forms of JMP and CALL are referred to as other segments and require privilege checking.

Hence, when the CPU fetches a far-call instruction, it will use that instruction’s ‘selector’ value to look up a descriptor in the GDT (or in the current LDT).
If the call-gate descriptor is fetched, and if access is allowed (i.e., if CPL $\leq$ DPL), then the CPU will perform a complex sequence of actions which will accomplish the requested ring-transition. CPL is based on the least significant 2-bits in register CS (also in SS).
\\
The new value for SS:SP comes from a special system-segment, known as the TSS (Task State Segment). The CPU locates its TSS by referring to the value in register TR (Task Register).

\subsection{Returning back to the user-mode}
After the call-gate is executed in kernel-mode, and we run shellcode in kernel-mode, it is time to return to the user-mode in order to avoid a crash in kernel-mode like BSOD in Windows or Kernel Panic in Linux.
\par In order to return to user-mode or any other outer ring that is used as the source of FAR CALL or FAR JMP, one should execute \textit{lret} instruction in the inner ring. It is analogous to the procedure when an interrupt is returned to the previous state.

\begin{enumerate}
\item Use the far-return instruction: ‘lret’
\begin{itemize}
\item Restores CS:IP from the current stack
\item Restores SS:SP from the current stack
\end{itemize}

\item Use the far-return instruction: ‘lret \$n’
\begin{itemize}

\item Restores CS:IP from the current stack
\item Discards n parameter-bytes from that stack
\item Restores SS:SP from that current stack
\end{itemize}
\end{enumerate}

\subsection{Combining attack with CWE-123}
CWE-123 stands for write-what-where bugs. We have employed CVE-2016-7255 to modify our specific GDT entries. Consequently, the kernel-mode code execution of the shell-code using a FAR CALL is achieved. Also, another effect of this attack is to change the supervisor bit of page table so that page tables are readable and writable in user-mode or \textit{self-ref of death attack)}. 

\section{Possible Mitigation}
The simple approach of complete isolation of the kernel is not able to fully unmap GDT from the user-mode since, in all modes of execution, the GDT descriptors should be available. 

Every segment register has a \textit{visible} part and a \textit{hidden} part. The hidden part sometimes referred to a \textit{descriptor cache} or a \textit{shadow register}. When a segment selector is loaded into the visible part of a segment register, the processor also loads the hidden part of the segment register with the base address, segment limit, and access control information from the segment descriptor pointed to by the segment selector. The information cached in the segment register (visible and hidden) allows the processor to translate addresses without taking extra bus cycles to read the base address and limit from the segment descriptor. In systems in which multiple processors have access to the same descriptor tables, it is the responsibility of software to reload the segment registers when the descriptor tables are modified. Otherwise, an old segment descriptor cached in a segment register might be used after its memory-resident version has been modified \cite{primary_procbased2019intel}.

It is worthy of mentioning that, complete mitigation to this attack would be the employment of separate GDT base in kernel and user layout. The kernel GDT should not be mapped into the user-mode, and Operating System Kernel has to change the address of GDTR each time a ring modification occurs. For example, it shall use SGDT to change the GDTR after every user-mode to kernel-mode switch caused by \textit{SYSENTER} and \textit{SYSCALL} or every interrupts handler routines.

The mapped GDT  in the user-mode should also be modifiable only by the kernel (not user-mode).

Hence, the user-mode application cannot access a valid address for GDT, and the discovered GDT address by the attacker is only valid when it is on user-mode. So, if a bug such as Write-What-Where occurs in the kernel or any system-level driver or kernel module, it cannot modify the user-mode GDT; thus, if the user-mode application tries to use call-gate in ring 3, the corresponding GDT entry is invalid, and the attack fails.


\section{Related Efforts}
Micro-architectural software attacks have been widely investigated in the context of revealing or damaging private and sensitive data. Recent works such as \cite{yarom2014flush,gras2018translation,van2018malicious} aim to discover data on the victim system secretly. Furthermore, adversary techniques for exploitation on shared Virtual Environments like \cite{oliverio2017secure} have shown to be promising in practice.

With regards to much older timing side-channel attacks, Osvik et al. \cite{osvik2006cache} introduced the PRIME+PROBE on the L1 cache, to attack the AES implementations, discovering secret keys. Consequently, more promising and sophisticated methods like \cite{yarom2014flush} were proposed.

Moreover, other software-based attacks on DRAM pioneered by \cite{kim2014flipping} have also shown to be very practical, jeopardizing the private data stored in memory systems in various circumstances.

In terms of exploiting the abandoned, but existing technologies in modern CPU designs, which is the primary concern of this paper, the possible vulnerabilities regarding the structure of GDT and IDT, were previously studied by \cite{jurczyk2010gdt}.
Researchers in \cite{jurczyk2010gdt} proposed a technique to gain a more stable kernel-level exploitation. These techniques were shown to be applicable in Windows-NT systems. Moreover, interestingly, several utilized mechanisms in this article, such as call-gate has also been used for securing the systems. For instance, \cite{lewis2013using} present an approach to prevent sandbox leakage based on call-gate.

\section{Conclusion and Discussion}

The impact of the hardware vulnerability exploited by software techniques has been proved to be dreadful. In this paper, we presented a TSX based side-channel attack, revealing the addresses of GDT and IDT in the kernel space, which could be exploited by an arbitrary user-mode application. We demonstrated that a single \textit{Write-What-Where}  vulnerability in the operating system could lead to a full system compromise through call-gate feature available in today's CPUs, irrespective of the version of the operating system. We have successfully evaluated our method by implementing an attack on the $9^{th}$ Generation \textit{Intel} processors.

The attack presented here is based on the descriptor structures available on the modern processors (e.g., Intel as well as AMD \cite{devices2006amd64}) although have hidden address by ASLR but are mapped into the user-mode address layout. The exploitation perfectly works with common \textit{Write What Where} bugs. For instance, any bug in a \textit{JavaScript} application on an isolated web-browser in the kernel address or graphic functions of the operating system (e.g., Win32k bugs in Windows) will be enough to be exploited. 
Moreover, we suggested software mitigation for this vulnerability since the presented attack bypasses the recent mitigation to Meltdown Attack (e.g., KAISER).

\section*{Acknowledgment}
The authors of this paper would like to thank the members of HPC lab in computer science department of IPM. We are also thankful specially to Mr. Kamyar Givaki for his consultants. Furthermore, we would like to thank Mr. Babak Amin Azad for his review and thoughtful feedback on the article. We're also very grateful to Petr Benes for his insights on the implementations on the Hyperviser. Moreover, we would like to send our regards to the contributors of \cite{monfared2019generating, daneshvaramoli2019decentralized} for their help.






\bibliographystyle{IEEEtran}
\bibliography{EuroSP}

\begin{thebibliography}{10}
\providecommand{\url}[1]{#1}
\csname url@samestyle\endcsname
\providecommand{\newblock}{\relax}
\providecommand{\bibinfo}[2]{#2}
\providecommand{\BIBentrySTDinterwordspacing}{\spaceskip=0pt\relax}
\providecommand{\BIBentryALTinterwordstretchfactor}{4}
\providecommand{\BIBentryALTinterwordspacing}{\spaceskip=\fontdimen2\font plus
\BIBentryALTinterwordstretchfactor\fontdimen3\font minus
  \fontdimen4\font\relax}
\providecommand{\BIBforeignlanguage}[2]{{%
\expandafter\ifx\csname l@#1\endcsname\relax
\typeout{** WARNING: IEEEtran.bst: No hyphenation pattern has been}%
\typeout{** loaded for the language `#1'. Using the pattern for}%
\typeout{** the default language instead.}%
\else
\language=\csname l@#1\endcsname
\fi
#2}}
\providecommand{\BIBdecl}{\relax}
\BIBdecl

\bibitem{8848043}
M.~S. {Kiarostami}, M.~{Reza Daneshvaramoli}, S.~K. {Monfared}, D.~{Rahmati},
  and S.~{Gorgin}, ``Multi-agent non-overlapping pathfinding with monte-carlo
  tree search,'' in \emph{2019 IEEE Conference on Games (CoG)}, 2019, pp. 1--4.

\bibitem{hajihassani2019fast}
O.~Hajihassani, S.~K. Monfared, S.~H. Khasteh, and S.~Gorgin, ``Fast aes
  implementation: A high-throughput bitsliced approach,'' \emph{IEEE
  Transactions on Parallel and Distributed Systems}, vol.~30, no.~10, pp.
  2211--2222, 2019.

\bibitem{ge2018survey}
Q.~Ge, Y.~Yarom, D.~Cock, and G.~Heiser, ``A survey of microarchitectural
  timing attacks and countermeasures on contemporary hardware,'' \emph{Journal
  of Cryptographic Engineering}, vol.~8, no.~1, pp. 1--27, 2018.

\bibitem{lipp2018meltdown}
M.~Lipp, M.~Schwarz, D.~Gruss, T.~Prescher, W.~Haas, A.~Fogh, J.~Horn,
  S.~Mangard, P.~Kocher, D.~Genkin \emph{et~al.}, ``Meltdown: Reading kernel
  memory from user space,'' in \emph{27th $\{$USENIX$\}$ Security Symposium
  ($\{$USENIX$\}$ Security 18)}, 2018, pp. 973--990.

\bibitem{kocher2018spectre}
P.~Kocher, D.~Genkin, D.~Gruss, W.~Haas, M.~Hamburg, M.~Lipp, S.~Mangard,
  T.~Prescher, M.~Schwarz, and Y.~Yarom, ``Spectre attacks: Exploiting
  speculative execution,'' \emph{arXiv preprint arXiv:1801.01203}, 2018.

\bibitem{weisse2018foreshadow}
O.~Weisse, J.~Van~Bulck, M.~Minkin, D.~Genkin, B.~Kasikci, F.~Piessens,
  M.~Silberstein, R.~Strackx, T.~F. Wenisch, and Y.~Yarom, ``Foreshadow-ng:
  Breaking the virtual memory abstraction with transient out-of-order
  execution,'' 2018.

\bibitem{minkin2019fallout}
M.~Minkin, D.~Moghimi, M.~Lipp, M.~Schwarz, J.~Van~Bulck, D.~Genkin, D.~Gruss,
  F.~Piessens, B.~Sunar, and Y.~Yarom, ``Fallout: Reading kernel writes from
  user space,'' \emph{arXiv preprint arXiv:1905.12701}, 2019.

\bibitem{schwarz2019zombieload}
M.~Schwarz, M.~Lipp, D.~Moghimi, J.~Van~Bulck, J.~Stecklina, T.~Prescher, and
  D.~Gruss, ``Zombieload: Cross-privilege-boundary data sampling,'' \emph{arXiv
  preprint arXiv:1905.05726}, 2019.

\bibitem{kurth_netcat}
\BIBentryALTinterwordspacing
M.~Kurth, B.~Gras, D.~Andriesse, C.~Giuffrida, H.~Bos, and K.~Razavi,
  ``{NetCAT}: {Practical} {Cache} {Attacks} from the {Network},'' in
  \emph{S\&{P}}, May 2020, intel Bounty Reward. [Online]. Available:
  \url{https://www.vusec.net/download/?t=papers/netcat_sp20.pdf}
\BIBentrySTDinterwordspacing

\bibitem{gruss2017kaslr}
D.~Gruss, M.~Lipp, M.~Schwarz, R.~Fellner, C.~Maurice, and S.~Mangard, ``Kaslr
  is dead: long live kaslr,'' in \emph{International Symposium on Engineering
  Secure Software and Systems}.\hskip 1em plus 0.5em minus 0.4em\relax
  Springer, 2017, pp. 161--176.

\bibitem{kim2014flipping}
Y.~Kim, R.~Daly, J.~Kim, C.~Fallin, J.~H. Lee, D.~Lee, C.~Wilkerson, K.~Lai,
  and O.~Mutlu, ``Flipping bits in memory without accessing them: An
  experimental study of dram disturbance errors,'' in \emph{ACM SIGARCH
  Computer Architecture News}, vol.~42, no.~3.\hskip 1em plus 0.5em minus
  0.4em\relax IEEE Press, 2014, pp. 361--372.

\bibitem{seaborn2015exploiting}
M.~Seaborn and T.~Dullien, ``Exploiting the dram rowhammer bug to gain kernel
  privileges,'' \emph{Black Hat}, vol.~15, 2015.

\bibitem{kvas2018}
\BIBentryALTinterwordspacing
S.~(Microsoft), ``Kva shadow: Mitigating meltdown on windows,'' 2018. [Online].
  Available:
  \url{https://msrc-blog.microsoft.com/2018/03/23/kva-shadow-mitigating-meltdown-on-windows/}
\BIBentrySTDinterwordspacing

\bibitem{grussKernelIsolation}
D.~Gruss, D.~Hansen, and B.~Gregg, ``Kernel isolation: From an academic idea to
  an efficient patch for every computer,'' \emph{; login: the USENIX Magazine},
  vol.~43, no.~4, pp. 10--14, 2018.

\bibitem{WestmereArrives}
\BIBentryALTinterwordspacing
D.~Kanter, ``Westmere arrives,'' 2010. [Online]. Available:
  \url{https://www.realworldtech.com/westmere/}
\BIBentrySTDinterwordspacing

\bibitem{TLB_and_Meltdown}
\BIBentryALTinterwordspacing
WikiZer0, ``Translation lookaside buffer,'' 2019. [Online]. Available:
  \url{https://www.wikizero.com/en/Translation_lookaside_buffer}
\BIBentrySTDinterwordspacing

\bibitem{IntelPCID}
\BIBentryALTinterwordspacing
I.~Corporation, ``4.10.1 process-context identifiers (pcids) - volume 3a:
  System programming guide, part 1,'' 2017. [Online]. Available:
  \url{https://software.intel.com/sites/default/files/managed/7c/f1/253668-sdm-vol-3a.pdf#page=139}
\BIBentrySTDinterwordspacing

\bibitem{GilTenePCID}
\BIBentryALTinterwordspacing
G.~Tene, ``Pcid is now a critical performance/security feature on x86,'' 2018.
  [Online]. Available:
  \url{https://groups.google.com/forum/m/#!topic/mechanical-sympathy/L9mHTbeQLNU}
\BIBentrySTDinterwordspacing

\bibitem{hvci}
\BIBentryALTinterwordspacing
S.~J. MICROSOFT, ``Virtualization based security (vbs) and hypervisor enforced
  code integrity (hvci) for olympia users!'' 2018. [Online]. Available:
  \url{https://techcommunity.microsoft.com/t5/Windows-Insider-Program/Virtualization-Based-Security-VBS-and-Hypervisor-Enforced-Code/m-p/240571}
\BIBentrySTDinterwordspacing

\bibitem{secondary_vmbased2019intel}
P.~Guide, ``Intel{\textregistered} 64 and ia-32 architectures software
  developer’s manual,'' \emph{Volume 3C: Chapter 24, VIRTUAL MACHINE CONTROL
  STRUCTURES (Table 24-7. Definitions of Secondary Processor-Based VM-Execution
  Controls)}, vol.~3C, 2019.

\bibitem{primary_procbased2019intel}
------, ``Intel{\textregistered} 64 and ia-32 architectures software
  developer’s manual,'' \emph{Volume 3C: Chapter 24, VIRTUAL MACHINE CONTROL
  STRUCTURES (Table 24-6. Definitions of Primary Processor-Based VM-Execution
  Controls)}, vol.~3C, 2019.

\bibitem{SinaKarvandi2018HVFS_Part1}
\BIBentryALTinterwordspacing
S.~Karvandi, ``Hypervisor from scratch – part 1: Basic concepts and configure
  testing environment,'' 2018. [Online]. Available:
  \url{https://rayanfam.com/topics/hypervisor-from-scratch-part-1/}
\BIBentrySTDinterwordspacing

\bibitem{SinaKarvandi2018HVFS_Part5}
\BIBentryALTinterwordspacing
------, ``Hypervisor from scratch – part 5: Setting up vmcs and running guest
  code,'' 2018. [Online]. Available:
  \url{https://rayanfam.com/topics/hypervisor-from-scratch-part-5/}
\BIBentrySTDinterwordspacing

\bibitem{uipi}
\BIBentryALTinterwordspacing
P.~katoch, ``Windows integrity mechanism design,'' 2016. [Online]. Available:
  \url{https://katochcisco.blogspot.com/2016/09/windows-integrity-mechanism-design.html}
\BIBentrySTDinterwordspacing

\bibitem{stecklina2018lazyfp}
J.~Stecklina and T.~Prescher, ``Lazyfp: Leaking fpu register state using
  microarchitectural side-channels,'' \emph{arXiv preprint arXiv:1806.07480},
  2018.

\bibitem{devices2006amd64}
A.~M. Devices, ``Amd64 architecture programmer’s manual volume 2: System
  programming,'' 2006.

\bibitem{SinaKarvandi2019CallGates}
\BIBentryALTinterwordspacing
S.~Karvandi, ``Call gates’ ring transitioning in ia-32 mode,'' 2019.
  [Online]. Available:
  \url{https://rayanfam.com/topics/call-gates-ring-transitioning-in-ia-32-mode/}
\BIBentrySTDinterwordspacing

\bibitem{ia32_gs_base2019intel}
P.~Guide, ``Intel{\textregistered} 64 and ia-32 architectures software
  developer’s manual,'' \emph{Volume 4: Chapter 2, MODEL-SPECIFIC REGISTERS
  (MSRS) (Table 2-2. IA-32 Architectural MSRs)}, vol.~4, 2019.

\bibitem{alex20181}
\BIBentryALTinterwordspacing
A.~Ionescu, ``blog post,'' 2018. [Online]. Available:
  \url{http://www.alex-ionescu.com/?p=340}
\BIBentrySTDinterwordspacing

\bibitem{cwe123}
\BIBentryALTinterwordspacing
MITRE, ``Cwe-123: Write-what-where condition,'' 2019. [Online]. Available:
  \url{https://cwe.mitre.org/data/definitions/123.html}
\BIBentrySTDinterwordspacing

\bibitem{inteltsd}
P.~Guide, ``Intel{\textregistered} 64 and ia-32 architectures software
  developer’s manual,'' \emph{Volume 3A: Chapter 1, SYSTEM ARCHITECTURE
  OVERVIEW, Time Stamp Disable )}, vol.~3A, 2019.

\bibitem{SinaKarvandi2019HVFS_Part6}
\BIBentryALTinterwordspacing
S.~Karvandi, ``Hypervisor from scratch – part 6: Virtualizing an already
  running system,'' 2019. [Online]. Available:
  \url{https://rayanfam.com/topics/hypervisor-from-scratch-part-6/}
\BIBentrySTDinterwordspacing

\bibitem{IntelFlexMigration}
\BIBentryALTinterwordspacing
Intel, ``Intel virtualization technology flexmigration application note,''
  2012. [Online]. Available:
  \url{https://www.intel.com/content/dam/www/public/us/en/documents/application-notes/virtualization-technology-flexmigration-application-note.pdf}
\BIBentrySTDinterwordspacing

\bibitem{schwarz2017malware}
M.~Schwarz, S.~Weiser, D.~Gruss, C.~Maurice, and S.~Mangard, ``Malware guard
  extension: Using sgx to conceal cache attacks,'' in \emph{International
  Conference on Detection of Intrusions and Malware, and Vulnerability
  Assessment}.\hskip 1em plus 0.5em minus 0.4em\relax Springer, 2017, pp.
  3--24.

\bibitem{schwarz2017fantastic}
M.~Schwarz, C.~Maurice, D.~Gruss, and S.~Mangard, ``Fantastic timers and where
  to find them: high-resolution microarchitectural attacks in javascript,'' in
  \emph{International Conference on Financial Cryptography and Data
  Security}.\hskip 1em plus 0.5em minus 0.4em\relax Springer, 2017, pp.
  247--267.

\bibitem{Osdev2017SysEnter}
\BIBentryALTinterwordspacing
O.~D. Wiki, ``Sysenter,'' 2017. [Online]. Available:
  \url{https://wiki.osdev.org/SYSENTER}
\BIBentrySTDinterwordspacing

\bibitem{yarom2014flush}
Y.~Yarom and K.~Falkner, ``Flush+ reload: a high resolution, low noise, l3
  cache side-channel attack,'' in \emph{23rd $\{$USENIX$\}$ Security Symposium
  ($\{$USENIX$\}$ Security 14)}, 2014, pp. 719--732.

\bibitem{gras2018translation}
B.~Gras, K.~Razavi, H.~Bos, and C.~Giuffrida, ``Translation leak-aside buffer:
  Defeating cache side-channel protections with $\{$TLB$\}$ attacks,'' in
  \emph{27th $\{$USENIX$\}$ Security Symposium ($\{$USENIX$\}$ Security 18)},
  2018, pp. 955--972.

\bibitem{van2018malicious}
S.~Van~Schaik, C.~Giuffrida, H.~Bos, and K.~Razavi, ``Malicious management
  unit: Why stopping cache attacks in software is harder than you think,'' in
  \emph{27th $\{$USENIX$\}$ Security Symposium ($\{$USENIX$\}$ Security 18)},
  2018, pp. 937--954.

\bibitem{oliverio2017secure}
M.~Oliverio, K.~Razavi, H.~Bos, and C.~Giuffrida, ``Secure page fusion with
  vusion: https://www. vusec. net/projects/vusion,'' in \emph{Proceedings of
  the 26th Symposium on Operating Systems Principles}, 2017, pp. 531--545.

\bibitem{osvik2006cache}
D.~A. Osvik, A.~Shamir, and E.~Tromer, ``Cache attacks and countermeasures: the
  case of aes,'' in \emph{Cryptographers’ track at the RSA conference}.\hskip
  1em plus 0.5em minus 0.4em\relax Springer, 2006, pp. 1--20.

\bibitem{jurczyk2010gdt}
M.~Jurczyk and G.~Coldwind, ``Gdt and ldt in windows kernel vulnerability
  exploitation,'' 2010.

\bibitem{lewis2013using}
P.~Lewis, ``Using a call gate to prevent secure sandbox leakage,'' Sep.~3 2013,
  uS Patent 8,528,083.

\bibitem{monfared2019generating}
S.~K. Monfared, O.~Hajihassani, S.~M. Zanjani, M.~Kiarostami, D.~Rahmati, and
  S.~Gorgin, ``Generating high quality random numbers: A high throughput
  parallel bitsliced approach,'' \emph{arXiv preprint arXiv:1909.04750}, 2019.

\bibitem{daneshvaramoli2019decentralized}
M.~Daneshvaramoli, M.~S. Kiarostami, S.~K. Monfared, H.~Karisani,
  H.~Khashehchi, D.~Rahmati, S.~Gorgin, and A.~Rahmati, ``Decentralized
  cooperative communication-less multi-agent task assignment with monte-carlo
  tree search,'' \emph{arXiv preprint arXiv:1910.12062}, 2019.

\end{thebibliography}


\end{document}